\documentclass{emulateapj}

\usepackage{amsmath}
\usepackage{amssymb}
\usepackage{graphicx}
\usepackage{color}
\usepackage{natbib}
\def\gtaprx {\lower .1ex\hbox{\rlap{\raise .6ex\hbox{\hskip .3ex
	{\ifmmode{\scriptscriptstyle >}\else
		{$\scriptscriptstyle >$}\fi}}}
	\kern -.4ex{\ifmmode{\scriptscriptstyle \sim}\else
		{$\scriptscriptstyle\sim$}\fi}}}
\def\ltaprx {\lower .1ex\hbox{\rlap{\raise .6ex\hbox{\hskip .3ex
	{\ifmmode{\scriptscriptstyle <}\else
		{$\scriptscriptstyle <$}\fi}}}
	\kern -.4ex{\ifmmode{\scriptscriptstyle \sim}\else
		{$\scriptscriptstyle\sim$}\fi}}}

\newcommand{\cutt}[1]{\textcolor{blue}{}}

\newcommand{\Ms}{{\ensuremath{M_{\odot} }}}

\begin{document}

\title{Radio Power from a Direct-Collapse Black Hole in CR7}

\author{Daniel J. Whalen\altaffilmark{1,2}, Mar Mezcua\altaffilmark{3,4}, Avery Meiksin\altaffilmark{5}, Tilman Hartwig\altaffilmark{6,7,8} and Muhammad A. Latif\altaffilmark{9}}

\altaffiltext{1}{Institute of Cosmology and Gravitation, Portsmouth University, Dennis Sciama Building, Portsmouth PO1 3FX}

\altaffiltext{2}{Ida Pfeifer Professor, Department of Astrophysics, University of Vienna, Tuerkenschanzstrasse 17, 1180, Vienna, Austria}

\altaffiltext{3}{Institute of Space Sciences (ICE, CSIC), Campus UAB, Carrer de Magrans, 08193 Barcelona, Spain}

\altaffiltext{4}{Institut d'Estudis Espacials de Catalunya (IEEC), Carrer Gran Capit\`{a}, 08034 Barcelona, Spain}

\altaffiltext{5}{Institute for Astronomy, University of Edinburgh, Blackford Hill, Edinburgh\ EH9\ 3HJ, UK}

\altaffiltext{6}{Kavli IPMU (WPI), UTIAS, The University of Tokyo, Kashiwa, Chiba 277-8583, Japan}

\altaffiltext{7}{Department of Physics, School of Science, University of Tokyo, Bunkyo, Tokyo 113-0033, Japan}

\altaffiltext{8}{Institute for Physics of Intelligence, School of Science, The University of Tokyo, Bunkyo, Tokyo 113-0033, Japan}

\altaffiltext{9}{Physics Department, College of Science, United Arab Emirates University, PO Box 15551, Al-Ain, UAE}

\begin{abstract}

The leading contenders for the seeds of the first quasars are direct collapse black holes (DCBHs) formed during catastrophic baryon collapse in atomically-cooled halos at $z \sim$ 20.  The discovery of the Ly$\alpha$ emitter CR7 at $z =$ 6.6 was initially held to be the first detection of a DCBH, although this interpretation has since been challenged on the grounds of {\em Spitzer} IRAC and Very Large Telescope X-Shooter data.  Here we determine if radio flux from a DCBH in CR7 could be detected and discriminated from competing sources of radio emission in the halo such as young supernovae and \ion{H}{2} regions.  We find that a DCBH would emit a flux of 10 - 200 nJy at 1.0 GHz, far greater than the sub-nJy signal expected for young supernovae but on par with continuum emission from star-forming regions. However, radio emission from a DCBH in CR7 could be distinguished from free-free emission from \ion{H}{2} regions by its spectral evolution with frequency and could be detected by the Square Kilometer Array in the coming decade.
 
\end{abstract}


\maketitle

\section{Introduction}

Over 300 quasars have now been discovered at $z >$ 6, including seven at $z >$ 7 \citep{mort11,ban18,mats19}.  The seeds of these quasars may be supermassive primordial stars that die as direct-collapse black holes (DCBHs) at $z \sim$ 20.  They form when atomic cooling in a 10$^7$ - 10$^8$ \Ms\ metal-free halo triggers catastrophic baryon collapse at its center, with infall rates of up to 1 \Ms\ yr$^{-1}$ \citep[e.g.,][]{rh09b,latif13a}.  These inflows build up a single star that later collapses to a BH at a mass of $\sim$ 10$^5$ \Ms\ \citep{hos13,tyr17,hle18,hle17}.  DCBHs are the leading candidates for the seeds of the first quasars because they are born in high accretion rates in which they grow more quickly than normal Pop III star BHs, which form in much lower densities \citep[e.g.,][]{wan04}.  Less massive Pop III star BHs are also subject to natal kicks that can eject them from their halos \citep{wf12} and do not encounter enough gas at later times to fuel their rapid growth (\citealt{srd18} -- see also \citealt{mez17,titans}).

The discovery of the strong Ly$\alpha$ emitter CR7 at $z =$ 6.6 \citep{bwl12} was originally held by some to be the first detection of a DCBH \citep[or a Pop III galaxy;][]{cr7} because of the detection of \ion{He}{2} 1640 \AA\ emission and the absence of metal lines in the initial observations.  Subsequent analyses favored a DCBH because of the difficulties associated with forming 10$^7$ \Ms\ of Pop III stars at the lower limit of metallicity imposed by observations at the time \citep{til15b}.  But this interpretation has since been challenged on the grounds of {\sc [OIII]} 4959 \AA\ and 5007 \AA\ emission in {\em Spitzer} IRAC data \citep{bwl17}, {\sc [CII]} 158~$\mu$m emission found by the Atacama Large Millimeter Array \citep[ALMA;][]{matth17}, and the re-analysis of Very Large Telescope (VLT) X-Shooter data \citep{shib18}, which failed to confirm the presence of \ion{He}{2} recombination line emission.  In particular, the presence of oxygen and carbon was thought to rule out a DCBH in CR7 because they form in zero-metallicity environments.

However, population synthesis and spectral fitting models predict masses of 5 - 10 million \Ms\ for a BH in CR7, well above those of DCBHs at birth, suggesting if one did form in CR7 it had since grown by up to a factor 100 in mass  \citep{agarw15b,pac17}.  If so, one would expect the existence of metals in CR7 because X-rays from the BH are known to trigger star formation and supernova (SN) explosions in its vicinity.  Secondary ionizations from energetic photoelectrons enhance free electron fractions and H$_2$ formation in the gas, which then cools and forms stars \citep[e.g.,][]{mba03}.  Metals or dust could also obscure \ion{He}{2} recombination line emission from the BH.

Radio observations could reveal the existence of a DCBH in CR7 because it could emit synchrotron radiation that could be detected by the next-generation Very Large Array (ngVLA) or epoch of reionization (EoR) observatories such as the Low Frequency Array (LOFAR) or the Square Kilometer Array (SKA).  Recent studies indicate that the amplification of seed magnetic fields by turbulent dynamos could create tangled magnetic fields even in primordial accretion disks \citep{schob12}.  These fields can then be ordered and further amplified by the rotation of the disk by the $\alpha$$\Omega$ dynamo \citep{ls15} and emit strong radio fluxes upon birth of the BH.  

However, young SN remnants in a starburst could masquerade as a DCBH by emitting large synchrotron fluxes at early times, and \ion{H}{2} regions due to star formation can also be sources of GHz emission \citep{reines20}.  The relative strengths of these three sources will determine if the detection of radio emission from CR7 would reveal the existence of a BH there.  We have calculated radio fluxes for a DCBH, SN remnants, and \ion{H}{2} regions in CR7.  In Section 2 we describe our empirical estimates of DCBH flux derived from several fundamental planes of BH accretion and calculations of radio fluxes due to young SN remnants and \ion{H}{2} regions.  We compare these fluxes in Section 3 to determine if the detection of radio emission from CR7 could indicate the presence of a BH.

\section{Radio Emission from CR7}

We consider radio flux from a DCBH, SNe, and \ion{H}{2} regions due to the formation of massive stars.

\subsection{DCBH}

Observations have empirically confirmed a correlation between the mass of a BH, $M_\mathrm{BH}$, its 2 - 10 keV nuclear X-ray luminosity, $L_\mathrm{X}$, and its 5 GHz nuclear luminosity, $L_\mathrm{R}$, known as the fundamental plane of BH accretion (\citealt{merl03}; see \citealt{mez18} for a brief review).  This correlation is supported by theoretical models of accretion and extends over six orders of magnitude in mass, including the intermediate mass black hole (IMBH) regime \citep{gul14}.  There has been some debate if radio-loud and radio-quiet active galactic nuclei (AGNs) occupy distinct regions in the FP but \citet{franca10} and \citet{bonchi13} have found no evidence for a bimodality in the radio luminosity function and that the FP is applicable to all types of AGNs.  

\begin{deluxetable}{lccc}
\tabletypesize{\scriptsize}
\tablecaption{Fundamental Planes\label{tbl:FPs}}
\tablehead{
\colhead{FP}  & \colhead{$\alpha$}  & \colhead{$\beta$}   & \colhead{$\gamma$}}
\startdata
MER03   &  0.60  & 0.78  &  7.33  \\
KOR06   &  0.71  &  0.62  &  3.55 \\
GUL09   &  0.67  &  0.78  &  4.80  \\
PLT12    &  0.69  &  0.61  &  4.19  \\
BON13   &  0.39  &  0.68  & 16.61 
\enddata
\end{deluxetable}

To estimate the flux from a DCBH in CR7 in a given radio band in the observer frame we first calculate its 5 GHz luminosity in the rest frame with a FP.  This requires $L_\mathrm{X}$, which we find from the bolometric luminosity of the BH with Equation 21 of \citet{marc04},
\begin{equation}
\mathrm{log}\left(\frac{L_\mathrm{bol}}{L_\mathrm{X}}\right) = 1.54 + 0.24 \mathcal{L} + 0.012 \mathcal{L}^2 - 0.0015 \mathcal{L}^3,
\end{equation}
where $\mathcal{L} = \mathrm{log} \, L_\mathrm{bol} - 12$ and $L_\mathrm{bol}$ is in units of solar luminosity.  \citet{agarw15b} estimate the mass and luminosity of the BH in CR7 to be 4.4 $\times$ 10$^6$ \Ms\ and 0.4 $L_\mathrm{Edd}$, respectively, where $L_\mathrm{Edd} =$ 1.26 $\times$ 10$^{38}$ ($M/$\Ms) erg s$^{-1}$. These values yield $L_\mathrm{X} =$ 1.22 $\times$ 10$^{43}$ erg s$^{-1}$, which is consistent with the upper limit $L_\mathrm{X} \lesssim$ 10$^{44}$ erg s$^{-1}$ found by \citet{pac17b}.  $L_\mathrm{R}$ can then be obtained from any of a number of FPs of the form
\begin{equation}
\mathrm{log} \, L_\mathrm{R} = \alpha \, \mathrm{log} \, L_\mathrm{X} + \beta \, \mathrm{log} \, M_\mathrm{BH} + \gamma,
\end{equation}
where $\alpha$, $\beta$ and $\gamma$ for FPs from \citet[][MER03]{merl03}, \citet[][KOR06]{kord06}, \citet[][GUL09]{gul09}, \citet[][PLT12]{plot12}, and \citet[][BON13]{bonchi13} are listed in Table~\ref{tbl:FPs}.  We also include the FP of Equation 19 of \citet[][GUL19]{gul19}, 
\begin{equation}
R \, = \, -0.62 + 0.70 \, X + 0.74 \, \mu,
\end{equation}
where $R =$ log($L_\mathrm{R}/10^{38} \mathrm{erg/s}$), $X =$ log($L_\mathrm{X}/10^{40} \mathrm{erg/s}$) and $\mu =$ log($M_\mathrm{BH}/10^{8}$\Ms). 

\begin{figure} 
\plotone{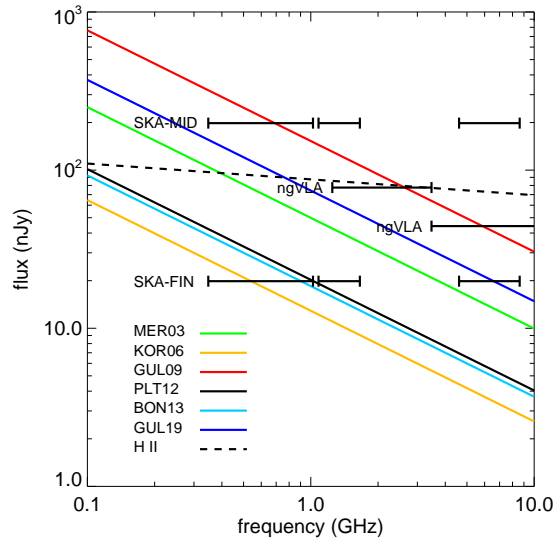}
\caption{CR7 DCBH (solid lines) and \ion{H}{2} fluxes (dashed line) predicted by the six FPs from 100 MHz - 10 GHz with detection limits for the SKA-MID and SKA-FINAL surveys and ngVLA sensitivities for 24 hr integration times.}
\vspace{0.1in}
\label{fig:flux} 
\end{figure}

Radio flux from CR7 that is redshifted into a given band in the observer frame in general does not originate from 5 GHz in the source frame so we calculate it from $L_\mathrm{R} =$ $\nu L_{\nu}$, assuming that the spectral luminosity $L_{\nu} \propto \nu^{-\alpha}$ with a spectral index $\alpha =$ 0.7 \citep{ccb02}.  The spectral flux at $\nu$ in the observer frame can then be obtained from the spectral luminosity at $\nu'$ in the rest frame from
\begin{equation}
F_\nu = \frac{L_{\nu'}(1 + z)}{4 \pi {d_\mathrm L}^2},
\end{equation}
where $d_\mathrm L$ is the luminosity distance and $\nu' = (1+z) \nu$.  If we use second-year \textit{Planck} cosmological parameters \citep[$\Omega_{\mathrm M} = 0.308$, $\Omega_\Lambda = 0.691$, $\Omega_{\mathrm b}h^2 = 0.0223$, $\sigma_8 =$ 0.816, $h = $ 0.677 and $n =$ 0.968;][]{planck2}, $d_\mathrm A =$ 1143.8 Mpc.  We plot DCBH fluxes from 100 MHz - 10 GHz for all six FPs in Figure~\ref{fig:flux}.

\subsection{SN Radio Flux}

\begin{figure} 
\epsscale{1.1}
\plotone{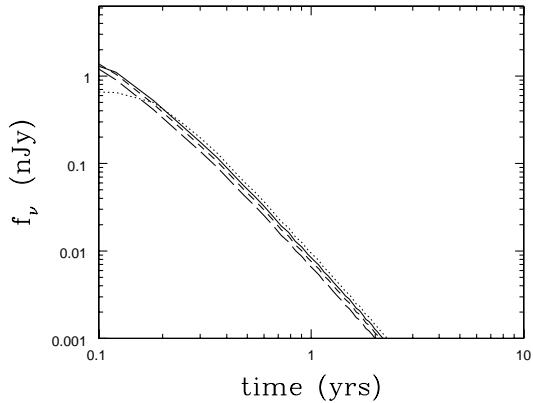}
\caption{Radio synchrotron emission from a 15 \Ms\ CC SN in a fiducial \ion{H}{2} region with an ambient density of 0.4 cm$^{-3}$ at $z =$ 6.6:  0.5 GHz (dotted), 1.4 GHz (solid), 3 GHz (short-dashed) and 8.4 GHz (long-dashed).
\label{fig:SN}}
\end{figure}

We first consider the simplest case of SNe due to a starburst.  In a starburst, most of the stars form in about the lifetime of any one of them, and 10 - 20 \Ms\ core-collapse (CC) SNe would produce the most synchrotron emission because energetic pair-instability (PI) SNe explode in much lower-density \ion{H}{2} regions that emit far less radio energy when swept up by the remnant \citep{mw12}.  Elemental abundances measured in a number of extremely metal-poor stars suggest that many stars in the early Universe may have been a few tens of solar masses \citep[e.g.,][]{jet09b,ish18}.  The SN rate in a starburst can be obtained by dividing the total mass of Pop III stars originally inferred to be in CR7, $\sim 10^7$ \Ms, by the average mass of the longest-lived stars capable of producing SNe, which we take to be $\sim$ 15 \Ms.  This yields the maximum number of SNe over the duration of the burst, which would be about the lifetime of a 15 \Ms\ Pop III star, $\sim$ 10 Myr.  Dividing the total number of SNe by the duration of the burst yields one Pop III SN every 15 yr.  Another SN rate can be derived from the Pop II star formation rate (SFR) inferred from recent observations of CR7 by ALMA, which is $\sim$ 50 \Ms\ yr$^{-1}$ \citep[Table 1 of][]{matth17}.  Assuming a Salpeter IMF with one CC SN per 60 \Ms\ of stars, this SFR produces about one CC SN per year.  

In Figure~\ref{fig:SN} we show radio fluxes for a single 15 \Ms\ CC SN at $z =$ 6.6 in densities like those expected for the low-metallicity environments of massive stars in CR7.  Ionizing UV flux from the progenitor star first creates an \ion{H}{2} region.  The star then explodes in ambient densities of 0.4~cm$^{-3}$ \citep[halo 2 of][]{wet08a}.  After the explosion the fluxes peak at $\sim$ 0.75 - 2 nJy but then fall by three orders of magnitude in 2 yr \citep[for a detailed description of this calculation, see][]{mw12}.  Although the SN peaks above 1 nJy, its average flux over the 15 yr interval between explosions in the Pop III starburst would be at least a factor of 10 lower because of its sharp decline in less than a year.  We have verified that these fluxes change very little as ambient densities are varied over an order of magnitude, which are typical of \ion{H}{2} regions in high-redshift, low-metallicity dwarf galaxies.  Even the Pop II SFR would yield less than 1 nJy of continuous flux, $\sim$ 10 - 1000 times less than that of a DCBH, depending on frequency.

\subsection{\ion{H}{2} Radio Flux}

Thermal bremsstrahlung in \ion{H}{2} regions can produce continuum radio emission whose spectral radio density can be connected to the ionizing photon rate in the \ion{H}{2} region, $Q_{\mathrm{Lyc}}$, by
\begin{equation}
L_{\nu} \, \lesssim \, \left(\frac{Q_{\mathrm{Lyc}}}{6.3 \times 10^{52}}\right) \left(\frac{T_{\mathrm{e}}}{10^4 \mathrm{K}}\right)^{0.45} \left(\frac{\nu}{\mathrm{GHz}}\right)^{-0.1}
\end{equation}
in units of 10$^{20}$ W Hz$^{-1}$ \citep{con92}, where $Q_{\mathrm{Lyc}} =$ SFR (\Ms yr$^{-1}$) $/$ $1.0 \times 10^{-53}$ \citep{ken98}.  We estimate the radio continuum from star-forming regions in CR7 by assuming SFR $=$ 50 \Ms\ yr$^{-1}$ and $T_{\mathrm{e}} =$ 10$^4$ K and plot it in Figure~\ref{fig:flux}.  It varies from 100 nJy at 100 MHz to 70 nJy at 10 GHz.  We take this flux to be an upper limit because it is calibrated for star-forming regions in the local Universe today.

\section{Discussion and Conclusion}

Our calculations indicate that radio emission from a DCBH in CR7 could be detected by the SKA and ngVLA.  The SKA-MID deep survey will reach sensitivities of 200 nJy in three bands (0.35 to 1.05 GHz, 0.95 to 1.76 GHz and 4.6 to 8.5 GHz) while the SKA-FINAL all-sky survey will reach 20 nJy in these bands\footnote{https://www.ectstar.eu/sites/www.ectstar.eu/files/talks/trento\_Wagg.pdf}.  The ngVLA could reach 45 nJy at 3.5 - 12.3 GHz and 78 nJy at 1.2 - 3.5 GHz in 24 hr integration times \citep{pr18}, which would be sufficient to detect the flux in those bands predicted by MER03, GUL09 and GUL19\footnote{http://library.nrao.edu/public/memos/ngvla/NGVLA\_21.pdf}.  No surveys for LOFAR would exceed sensitivities of a few $\mu$Jy so it is unlikely to detect any radio emission from CR7.  The VLA has already visited the COSMOS legacy fields in which CR7 was originally discovered at 3 GHz with a sensitivity of 2.3 $\mu$Jy/beam \citep{smol17} but we found no radio counterpart to CR7 in this archive (none of the FPs predict fluxes of this magnitude).

It is unlikely that radio emission from SNe in CR7 could be mistaken for DCBH emission because their average fluxes would be less than a nJy, even from starbursts.  This  signal would be well below the detection limit of any planned survey and is a factor of 10 - 100 smaller than even the most pessimistic DCBH fluxes predicted by the FPs.  SN radio luminosities compiled for a sample of 19 nearby galaxies \citep{cw09} predict higher average fluxes for the SFRs inferred for CR7 \citep[8.5 - 85 nJy from 10 GHz to 100 MHz; Section 5.3 of][]{reines20}.  If these values were true of SN populations in CR7 they could still be distinguished from emission from a DCBH because they are lower than all but two of the fluxes predicted by the FPs.  However, it is unlikely that SN remnants in CR7 would emit this much flux.  Absorption of ionizing UV by dust in \ion{H}{2} regions at solar metallicities today limits them to smaller radii and thus higher densities, and stellar winds also plow up ambient gas and create dense structures in the vicinity of the stars.  Ejecta from SNe crashing into these higher densities emit considerably more radio flux than SNe in the diffuse \ion{H}{2} regions of low-metallicity environments at high redshift, in which stellar winds are weak if present at all \citep{wan04}.

The continuum radio flux due to \ion{H}{2} regions in CR7 could be similar to or even greater than that of a DCBH depending on frequency and choice of FP.  However, this flux falls off much more slowly with frequency than DCBH emission so the two could be easily distinguished at frequencies below about 3 GHz by the SKA for half of the FPs.  Radio emission due to thermal bremsstrahlung ultimately depends on electron temperatures and densities.  Because our estimates here are derived for \ion{H}{2} regions in local galaxies, which have higher densities than those in less-massive, high-redshift galaxies, we take them to be a (possibly severe) upper limit to the \ion{H}{2} region radio flux from CR7.  If the true flux is considerably smaller then ngVLA could still find emission from a DCBH.  Otherwise, the detection of a DCBH in CR7 due to disparities in flux from \ion{H}{2} regions as a function of frequency may be limited to the SKA.
  
A unique aspect of high-redshift quasars is that the cosmic microwave background (CMB) can quench radio emission from BH jets.  If the energy density of CMB photons exceeds that of the  magnetic fields in the lobes of the jet, relativistic electrons preferentially cool by upscattering CMB photons rather than synchrotron radiation, and the lack of radio emission from some high-redshift quasars has been attributed to this process \citep{gh14,fg14}.  However, this would not change the fluxes in our calculations because they come from the central region of the quasar, not jets, and jets are not expected at the accretion rates estimated for the BH in CR7 because they have only been observed at $L \lesssim 0.01 \, L_{\mathrm{Edd}}$ and $L \gtrsim L_{\mathrm{Edd}}$.  Finally, we note that while the detection of radio emission could confirm the presence of a BH in CR7, the failure to do so would not rule out its existence.  It could be that the radio fluxes associated with DCBH candidates lie below those predicted by FPs today so the discovery of a BH in CR7 may have to await future observatories.

\acknowledgments

The authors thank the anonymous referee for constructive comments that improved the quality of this paper, and Bhaskar Agarwal, Philip Best, Simon Glover and Marta Volonteri for helpful discussions.  D. J. W. was supported by STFC New Applicant Grant ST/P000509/1 and the Ida Pfeiffer Professorship at the Institute of Astrophysics at the University of Vienna.  M. M. acknowledges support from the Beatriu de Pinos fellowship (2017-BP-00114).  A. M. acknowledges support from the UK Science and Technology Facilities Council Consolidated Grant ST/R000972/1.  T. H. was supported by JSPS KAKENHI Grant Number 17F17320.  M. L. acknowledges funding from UAEU via UPAR grant No. 31S372.

\bibliographystyle{apj}
\bibliography{refs}

\end{document}